\documentclass[preprint, aps, pre, amsmath,amssymb, superscriptaddress]{revtex4-1}

\usepackage{graphicx}
\usepackage{dcolumn}
\usepackage{bm}
\usepackage{color}
\usepackage{enumitem}
\usepackage[normalem]{ulem}
\usepackage{comment}

\newif\ifHighlitedChanges
\def\ifHighlitedChanges{\iffalse}
\ifHighlitedChanges
  \def\EDITS#1{{\color{red}#1}}
  \def\STRIKE#1{{\color{red}\sout{#1}}}
\else
  \def\EDITS#1{#1}
  \def\STRIKE#1{\relax}
\fi

\begin{document}
\title{Heat Transport Hysteresis Generated through Frequency Switching of a Time-Dependent Temperature Gradient}
\author{Renai Chen}
\affiliation{Theoretical Division and Center for Nonlinear Studies, Los Alamos National Laboratory, Los Alamos, New Mexico, USA}
\author{Galen T. Craven}
\affiliation{Theoretical Division, Los Alamos National Laboratory, Los Alamos, New Mexico, USA}
\begin{abstract}A stochastic energetics framework is applied to examine how periodically shifting the frequency of a time-dependent oscillating temperature gradient
affects heat transport in a nanoscale molecular model.
We specifically examine the effects that frequency switching, i.e., instantaneously changing the oscillation frequency of
the temperature gradient, has on the shape of the heat transport hysteresis curves generated by a particle connected to two thermal baths,
each with a temperature that is oscillating in time.
Analytical expressions are derived for the energy fluxes in/out of the system and the baths, 
with excellent agreement observed between the analytical expressions and the results from nonequilibrium molecular dynamics simulations.
We find that the shape of the heat transport hysteresis curves
can be significantly altered by shifting the frequency between fast and slow oscillation regimes.
We also observe the emergence of features in the hysteresis curves such as pinched loops and complex multi-loop patterns due to the frequency shifting.
The presented results have implications in the design of thermal neuromorphic devices such as thermal memristors and thermal memcapacitors.
\end{abstract}

\maketitle

\section{Introduction}
Heat transport hysteresis is a phenomenon in which an energy flux generated by a temperature gradient that is oscillating in time is out of phase with the  oscillation pattern of the gradient itself.
It results in dynamic lagging between the driving force (the oscillating temperature gradient) and the system output (the energy flux).
This out-of-phase transport behavior gives rise to memory effects in the heat transport processes
that can be used, for example, for storing energy and information.
Heat transport hysteresis has been observed in macroscale systems \cite{Ben-Abdallah2017thermalmemristor,Ordonez-Miranda2019thermalmemristor} and
microscale systems \cite{craven2023b}.
However, many concepts from dynamical systems theory and nanoscale device design such as frequency modulation, frequency switching, and piecewise smooth
forcing, are currently underexplored in the context of molecular heat transport hysteresis.
The effects of these concepts on nanoscale heat transport and their ramifications for technological advancement are therefore currently unknown.

Nanoscale energy transport is a fundamental physical phenomenon that has received significant research interest for decades
\cite{Lebowitz1959,Lebowitz1967,Lebowitz1971,Cahill2003,Segal2005prl,Segal2016,Nitzan2007,Sato2012,Maldovan2013,Leitner2008,Leitner2013,Seifert2015periodictemp,Li2012,Dubi2011,Lim2013,craven16c,He2021,Volz2022, HernandezJPCL2023, sharony2020stochastic, Krivtsov2023,Dastgeer2022,Dastgeer2024}. 
At the molecular level, i.e, the microscopic level, the principles that are used to understand heat transport at the macroscale such as Fourier's law
are generally not applicable \cite{craven2024a}. 
This is because of several factors including the complex coupling between multiple heat transport mechanisms that is common at the nanoscale.
Specifically, in nanoscale systems, phononic and electronic heat transport mechanisms, as well as mechanisms that include coupling between them,
can be present and influence heat transport \cite{craven2023a,Segal2005jcp,Wu2007,Segal2008prl,Segal2009,Leitner2013,craven18b,Simon2021}.
Furthermore, in these mechanisms the transport processes can be ballistic or diffusive, or exist in some intermediate regime between the two,
and this further compounds the complexity and richness of nanoscale heat transport.
Interesting transport phenomenon such as thermoelectric rectification \cite{Kuo2010, craven18b} and thermal rectification \cite{Li2004,He2006,Wu2007,Kuo2010,Roberts2011,Giazotto2015,Zhao2023,RomeroBastida2024}
can be used design nanoscale devices that utilize temperature gradients and waste heat for useful operations \cite{Li2004,Wu2007,Kuo2010,Roberts2011,Giazotto2015,Zhao2023,craven2023a,Segal2005jcp,Segal2008prl,Segal2009,Leitner2013,craven18b,Simon2021,Krivtsov2023}.
Moreover, controlling and harnessing heat generated by a process for useful operations has implications in the design of novel thermal computing devices that use heat instead of electricity to perform computations \cite{Li2006,Ben-Abdallah2014,Joulain2016,Wang2017thermaldiode,craven17a,craven18b, Odebowale2024}.

Stochastic thermodynamics is a theoretical framework that can be applied to describe energetic properties 
of small systems such as single molecules when they are driven far from equilibrium \cite{Seifert2012,Van2013stochastic}.
In the context of examining systems with time-dependent temperatures and/or systems that are subjected to a time-dependent temperature gradient, a number of theoretical formalisms have been developed to understand the transport processes that emerge in such systems \cite{Reimann2002,brey1990generalized,hern07a,hern13d,Ford2015,Seifert2015periodictemp,Seifert2016periodiccurrent,Awasthi2021,Portugal2022effective,Volz2022,Weron2022,Ben-Abdallah2017thermalmemristor,Ordonez-Miranda2019thermalmemristor,Krivtsov2020}.
We have previously developed theories that utilize stochastic interpretations of energy transport processes to describe heat transport through nanoscale systems and molecular lattice structures when they are subjected to a temperature gradient that is oscillating in time \cite{craven2023b, craven2024a, Chen2024}.
In the context of developing thermal computing devices such as thermal memristors and memcapacitors that utilize heat transport hysteresis to enable their functionality, stochastic energetics formalisms allow the modeling of the energy transport and memory effects critical for designing neuromorphic functionalities. 

In dynamical systems theory, frequency switching is the process of altering the oscillation frequency of a  driving force, external field, or system parameters. The switching is often performed periodically or in an instantaneous manner when the system reaches a specific boundary, such as when a system trajectory reaches a specific region in phase space \cite{bernardo2008piecewise}. This technique can be used to elucidate how transitions between different frequency regimes impact properties such as a system's stability or propensity to promote energy transport. Frequency switching can give rise to complex phenomena such as bifurcations and hysteresis effects, and can be used to understand how a system will evolve in time \cite{Bonet2021, bonet2023novel, han2023sliding, jiang2024novel}. For example, frequency switching can be use to determine the paths a system will take to transition between two different equilibrium states when  conditions or parameters change, often abruptly.

In this article, we examine how an oscillating temperature gradient undergoing frequency switching influences heat transport properties, particularly the shape of heat transport hysteresis curves, in nanoscale systems. 
The study focuses on a model in which a single particle interacts with two thermal baths, where each bath has a temperature that oscillates in time with one 
of the bath temperatures undergoing frequency shifting.
A Langevin equation of motion coupled with a stochastic thermodynamics description of the energetic properties and energy fluxes is used to describe the thermodynamics of the model. Analytical solutions for the energy fluxes in/out of the system and the baths are derived and validated against stochastic molecular dynamics simulations.
A specific focus is placed on elucidating how periodically altering the frequency of the time-dependent temperature gradient impacts heat transport dynamics and hysteresis.
The effects of frequency switching results in complex hysteresis curve patterns.  We illustrate how frequency shifting of a temperature gradient generates emergent features in the heat transport hysteresis curves, including the formation of pinched loops and intricate multi-loop structures.

The rest of the article is organized as follows:
Section~\ref{sec:model} contains the details of the model used to 
examine the effects that a frequency-switched oscillating temperature gradient has 
on the heat transport properties of a paradigmatic nanoscale model.
Section~\ref{sec:flux} contains derivations of the time-dependent energy flux expressions
in the model. 
In Section~\ref{sec:hys}, we apply the derived energy flux expressions to examine
how frequency switching can be used to modify and influence 
heat transport hysteresis curves.
In the final section, conclusions and a discussion of potential future work are given.

\section{Model Details\label{sec:model}}

\begin{figure}[]
\includegraphics[width = 9.5cm,clip]{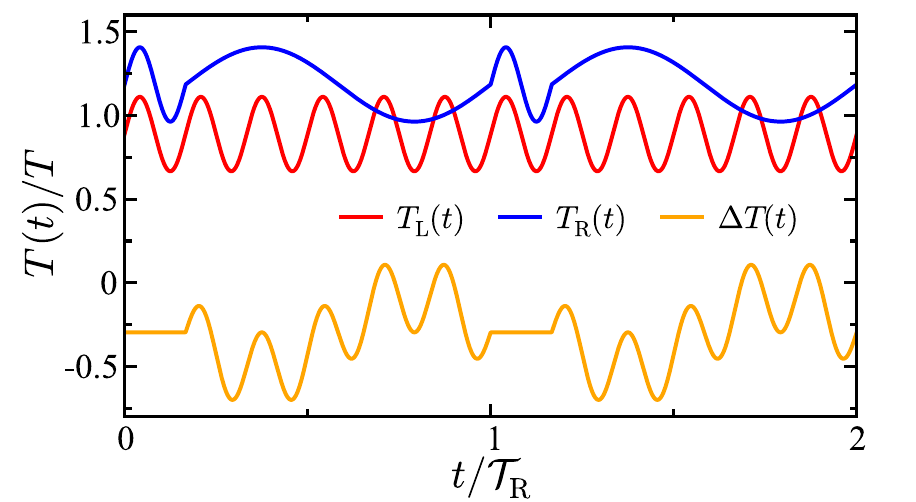}
\caption{\label{fig:temp}
Time-dependence of the left and right bath temperatures and the temperature difference between them.
The temperatures are shown in units of $T$ and time is shown in units of  $\mathcal{T}_\text{R}$.
Parameters are $\gamma = 2$ ($\gamma_\text{L} = 1$, $\gamma_\text{R} = 1$), $T^{(0)}_\text{L} = 4/5$, $T^{(0)}_\text{R} = 16/15$,
$\Delta T_\text{L} = 0.2$, $\Delta T_\text{R} = 0.2$, $\omega_\text{L} =5$, $\omega_{\text{R}_1} = 5$, $\omega_{\text{R}_2} = 1$.
All parameters throughout the article are given in reduced units with 
characteristic dimensions: $\widetilde{\sigma} = 1\,\text{\AA}$,  $\widetilde{\tau} = 1\,\text{ps}$,
$\widetilde{m} = 10\,m_u$,
and $\widetilde{T} = 300\,\text{K}$.
}
\end{figure}

The model we use consists of a single particle that is in contact with two heat baths, denoted ``L" for left bath and ``R" for right bath, both of which 
have time-dependent temperatures $T_\text{L}(t)$ and $T_\text{R}(t)$ that are oscillating in time. The temperature of the left bath oscillates at a constant single frequency $\omega_\text{L}$ while the temperature of the right bath is oscillating with a time-dependent frequency $\omega_\text{R}(t)$ that is being switched between two frequencies $\omega_{\text{R}_1}$ and $\omega_{\text{R}_2}$ in a periodic pattern. Therefore, the right bath is undergoing frequency switching in time. Further details on the specific forms for the 
temperatures are given later. 
The stochastic Langevin equation of motion for the system \EDITS{in the classical limit} is
\begin{equation}
\begin{aligned}
\label{eq:EoM1}
\dot x &= v, \\
\dot v &= - \gamma_\text{L} \dot x  - \gamma_\text{R} \dot x - m^{-1}\partial_x U(x) +  \xi_\text{L}(t) + \xi_\text{R}(t), 
\end{aligned}
\end{equation}
where $x$ is the position of the particle, $\dot{x} = v$ is the particle velocity, $\gamma_\text{L}$ and $\gamma_\text{L}$ parameterize the system-bath coupling strength for the baths,
$U(x)$ is the potential energy, 
$\xi_\text{L}(t)$ and $\xi_\text{R}(t)$ are stochastic noise terms, and $m$ is the particle mass.
We will refer to the particle as the ``system".
In this work, we will examine two cases for the potential. First, a free particle with $U(x) =0$ and, second, a particle in a harmonic potential $U(x) = \tfrac{1}{2}mkx^2$.
The noise terms obey the following correlations and cross-correlation:
\begin{equation}
\begin{aligned}
\label{eq:noise}
 \big\langle \xi_\text{L}(t) \xi_\text{L}(t')\big\rangle &= 2 \gamma_\text{L} k_\text{B} m^{-1} T_\text{L}(t)\delta(t-t'), \\[0ex]
 \big\langle \xi_\text{R}(t) \xi_\text{R}(t')\big\rangle &= 2 \gamma_\text{R} k_\text{B} m^{-1} T_\text{R}(t)\delta(t-t'), \\[0ex]
 \big\langle \xi_\text{L}(t)\big\rangle &=0, \\[0ex]
 \big\langle \xi_\text{R}(t)\big\rangle &=0, \\[0ex]
 \big\langle \xi_\text{L}(t) \xi_\text{R}(t')\big\rangle &= 0,  
\end{aligned}
\end{equation}
where $k_\text{B}$ is the Boltzmann constant.
The temperatures are periodic functions:
$T_\text{L}(t) = T_\text{L}(t+ \mathcal{T}_\text{L})$ and $T_\text{R}(t) = T_\text{R}(t+ \mathcal{T}_{\text{R}_1}+ \mathcal{T}_{\text{R}_2})$ 
where $\mathcal{T}_\text{L} = 2 \pi / \omega_\text{L}$ is the period of the left bath temperature oscillation and $\mathcal{T}_{\text{R}_1} = 2 \pi / \omega_{\text{R}_1}$ and  $\mathcal{T}_{\text{R}_2} = 2 \pi / \omega_{\text{R}_2}$ are the respective periods of oscillation for the right bath temperature
at each of the two frequencies.
The overall period for the right bath is $\mathcal{T}_{\text{R}} = \mathcal{T}_{\text{R}_1} +  \mathcal{T}_{\text{R}_2}$.
The time in the period of the right bath is defined by function:
\begin{equation}
t_\text{mod} = t\,(\text{mod}\, \mathcal{T}_{\text{R}}),
\end{equation}
The right bath is undergoing frequency switching
with the time-dependence of the switching frequency being formally defined as: 
\begin{equation}
\omega_\text{R}(t)  = 
    \begin{cases}
       \omega_{\text{R}_1} , & 0 \leq t_\text{mod} < \mathcal{T}_{\text{R}_1},\\[1ex]
       \omega_{\text{R}_2}, &   \mathcal{T}_{\text{R}_1}  \leq t_\text{mod} < \mathcal{T}_{\text{R}_2}.
    \end{cases} 
\end{equation}
We consider the case in which the temperatures of each bath take the specific forms
\begin{widetext}
\begin{align}
 T_\text{L}(t)   &= T^{(0)}_\text{L} + \Delta T_\text{L} \sin(\omega_\text{L} t),\\[1ex]
 T_\text{R}(t)   &=  
 \begin{cases}  
 T^{(0)}_\text{R}  +  \Delta T_\text{R} \sin(\omega_{\text{R}_1} t_{\text{mod}}) , & 0 \leq t_{\text{mod}} < \mathcal{T}_{\text{R}_1},\\ 
 T^{(0)}_\text{R} +   \Delta T_\text{R} \sin\left(\omega_{\text{R}_2} t_{\text{mod}} + \phi\right) , &\mathcal{T}_{\text{R}_1}  \leq t_{\text{mod}} < \mathcal{T}_{\text{R}_1} +  \mathcal{T}_{\text{R}_2}, \end{cases}
\end{align}
\end{widetext}
with $\phi = \mathcal{T}_{\text{R}_1}(\omega_{\text{R}_1}  - \omega_{\text{R}_2})$
being a phase factor that makes the $T_\text{R}(t)$ function continuous at the point the frequency is switched,
$T^{(0)}_\text{L}$ and $T^{(0)}_\text{R}$ are the temperatures of the two baths in the limit of vanishing of oscillations and $\Delta T_\text{L}$ and $\Delta T_\text{R}$ define the amplitude of the oscillations.
The instantaneous temperature difference between the two baths is 
\begin{equation}
\Delta T(t) =    T_\text{L}(t) - T_\text{R}(t).
\end{equation}
In the limit that the oscillations of both baths vanish, the effective temperature of the system is
\begin{equation}
\label{eq:temp}
T = \frac{\gamma_\text{L} T^{(0)}_\text{L} + \gamma_\text{R} T^{(0)}_\text{R}}{\gamma},
\end{equation}
with 
\begin{equation}
\gamma = \gamma_\text{L}+\gamma_\text{R}.
\end{equation}
Figure~\ref{fig:temp} shows an example of the time-dependence of the temperatures of each bath. Observe  
that the right bath temperature shown in blue has a frequency that switches in a oscillatory pattern.
\EDITS{At $t = 0$, the temperature is oscillating at frequency  $\omega_{\text{R}_1}>\omega_{\text{R}_2}$. Then, after one period of the higher frequency oscillation,
the frequency switches to $\omega_{\text{R}_2}$ and begins to oscillate at that lower frequency. 
At $t = \mathcal{T}_{\text{R}_1}+ \mathcal{T}_{\text{R}_2}$ (in the plot corresponding to $t = 1$  on the $x$-axis in scaled units),
the frequency switches back to the higher frequency oscillation, and the cycle of frequency switching continues.}
Frequency switching can be used to induce exotic transport phenomenon and to go between different transport regimes such as
switching between quasistatic and highly-inertial regimes of transport.

To describe the thermodynamic properties of the system we use the Sekimoto formalism for stochastic energetics \cite{Sekimoto1998}.
The Sekimoto formalism involves separating the terms in Langevin equation in Eq.~(\ref{eq:EoM1}) into two terms: terms that represent energy change in the system (the particle) and terms that contribute to the energy change in the baths. 
\EDITS{A stochastic energetics framework is used to examine the energy fluxes because of the small system size, the need to use a formalism that can account for the  system being driven far from equilibrium by the oscillating temperatures, and to build on previous work \cite{craven18a1, craven18a2, BaratoPRL2018}.}
As we have done in previous work \cite{craven2023b, craven2024a, Chen2024}, 
the expected energy fluxes in our model are separated into  three terms:
\begin{enumerate}
  \item $J_\text{sys}$ is the energy flux in/out of the system
  \item $J_\text{L}$ is the energy flux associated with the left bath 
  \item $J_\text{R}$ is the energy flux associated with the right bath
\end{enumerate}
A sum over all the energy fluxes obeys conservation of energy:
\begin{equation}
J_\text{L} (t)+J_\text{R} (t) + J_\text{sys} (t) = 0.
\end{equation}
The expectation values for the energy fluxes can be expressed as \cite{craven2023b,Lebowitz1959,Sekimoto1998,Sabhapandit2012,Dhar2015}:
\begin{align}
\label{eq:heatcurrentbathL}
J_\text{L} (t) &=  m \gamma_\text{L} \big\langle  v^2(t) \big\rangle-m \big\langle \xi_\text{L}(t) v(t)\big\rangle ,\\[1ex]
\label{eq:heatcurrentbathR}
J_\text{R} (t) &=  m \gamma_\text{R} \big\langle  v^2(t) \big\rangle-m \big\langle \xi_\text{R}(t) v(t)\big\rangle ,\\[1ex]
\label{eq:heatcurrentbathS}
J_\text{sys} (t) &= \frac{d \big\langle E(t)\big\rangle}{dt},
\end{align}
where $E(t)$ is the energy of the system, $\big\langle  v^2(t) \big\rangle$ is a velocity correlation function, and $\big\langle \xi_\text{L}(t) v(t)\big\rangle$ and $\big\langle \xi_\text{R}(t) v(t)\big\rangle$ are noise-velocity correlation functions.
Therefore, in order to derive expressions for the energy fluxes we must evaluate these time correlation functions.
Because the bath temperatures are varying, the system approaches a time-dependent nonequilibrium state (TDNES).
Note that if the system approached a nonequilibrium steady state, the time derivative of the system energy, i.e., the system energy flux, would go to zero. 
But, because of the oscillating temperature gradient, the system energy flux does not go to zero.

\section{Derivation of Energy Flux Expressions\label{sec:flux}}

To derive the heat transport properties, we need to calculate the correlation functions in Eqs.~(\ref{eq:heatcurrentbathL}) and (\ref{eq:heatcurrentbathR}).
We have previously examined the case without frequency switching in Ref.~\citenum{craven2023b}, and that is the starting point for the current derivation.

\subsection{Free Particle}
We first examine the case of a free particle with $U(x) = 0$.
The formal solutions of the equations of motion are

\begin{equation}
\begin{aligned}
x(t) &= x_0 + \int_0^t v(s) \,ds, \\[1ex]
\label{eq:vsol}
v(t) &=  v_0 e^{- \gamma t}   + \int_0^t  e^{-\gamma(t-s)}\xi_\text{L}(s)\,ds  + \int_0^t  e^{-\gamma(t-s)}\xi_\text{R}(s)\,ds,
\end{aligned}
\end{equation}
where $x_0$ is the initial position and $v_0$ is the initial velocity of the particle and $\gamma = \gamma_\text{L} + \gamma_\text{R}$.
Using the formal solutions, the noise-velocity correlation functions for each bath can be evaluated yielding
\begin{widetext}
\begin{align}
\label{eq:velnoisecorrL}
\nonumber\big\langle \xi_\text{L}(t) v(t)\big\rangle &= 
\big\langle \xi_\text{L}(t) v_0 \big\rangle  e^{-\gamma t}
+\int_0^t e^{-\gamma (t-s)} \big\langle\xi_\text{L}(t)\xi_\text{L}(s)\big\rangle \,ds +\int_0^t e^{-\gamma (t-s)} \big\langle\xi_\text{L}(t)\xi_\text{R}(s)\big\rangle \,ds \\
& = \frac{\gamma_\text{L} k_\text{B} T_\text{L}(t)}{m},\\
\label{eq:velnoisecorr}
\nonumber\big\langle \xi_\text{R}(t) v(t)\big\rangle &= 
\big\langle \xi_\text{R}(t) v_0 \big\rangle  e^{-\gamma t}
+\int_0^t e^{-\gamma (t-s)} \big\langle\xi_\text{R}(t)\xi_\text{L}(s)\big\rangle \,ds +\int_0^t e^{-\gamma (t-s)} \big\langle\xi_\text{R}(t)\xi_\text{R}(s)\big\rangle \,ds \\
& = \frac{\gamma_\text{R} k_\text{B} T_\text{R}(t)}{m}.
\end{align} 
\end{widetext}
The velocity correlation function is
\begin{equation}
\begin{aligned}
\label{eq:vsqrBrownian}
\big\langle v^2(t)\big\rangle  &=  
 v^2_0 e^{- 2\gamma t}
 + \int_0^t\!\!\int_0^t  e^{-\gamma(2t-s_1-s_2)}\big\langle\xi_\text{L}(s_1)\xi_\text{L}(s_2)\big\rangle\,ds_1\,ds_2 \\[1ex]
&\quad + \int_0^t\!\!\int_0^t  e^{-\gamma(2t-s_1-s_2)}\big\langle\xi_\text{R}(s_1)\xi_\text{R}(s_2)\big\rangle\,ds_1\,ds_2\\[1ex]
& \quad + 2\!\int_0^t e^{-\gamma(2 t-s_1)}\big\langle  \xi_\text{L}(s_1)v_0\big\rangle\,ds_1
+2 \!\int_0^t e^{-\gamma(2t-s_1)}\big\langle  \xi_\text{R}(s_1)v_0 \big\rangle\,ds_1 \\[1ex]
&\quad + 2\!\int_0^t\!\!\int_0^t  e^{-\gamma(2t-s_1-s_2)}\big\langle\xi_\text{L}(s_1)\xi_\text{R}(s_2)\big\rangle\,ds_1\,ds_2,\\[1ex]
&= v^2_0 e^{-2 \gamma t}
+ \frac{k_\text{B}T}{m} \left(1-e^{-2 \gamma t}\right) + \frac{2 k_\text{B}}{m}\bigg(\frac{ \gamma_\text{L} \Delta T_\text{L} (2 \gamma \sin(\omega_\text{L} t) - \omega_\text{L} \cos(\omega_\text{L} t)+ \omega_\text{L} e^{-2 \gamma t})}{4 \gamma^2+\omega^2_\text{L}} \\
& \quad+ e^{-2 \gamma t} \gamma_\text{R} \Delta T_\text{R} I(t,\gamma) \bigg),
\end{aligned}
\end{equation}
where
\begin{equation}
I(t,\gamma) = I_1 (t,\gamma) + I_2 (t,\gamma) + I_3 (t,\gamma),
\end{equation}
is a piecewise integral that is broken into three parts, each representing a different regime in the time interval $[0,t]$.
The first term, $I_1 (t,\gamma)$, accounts for time segments in completed right bath oscillation periods $\mathcal{T}_\text{R}$  when the temperature is oscillating at frequency $\omega_{\text{R}_1}$, and can be expressed as 
\begin{equation}
I_1 (t,\gamma) = \sum _{n=0}^{\left\lfloor t/\mathcal{T}_\text{R}\right\rfloor -1 } f_1\big((n+1) \mathcal{T}_{\text{R}_1}+ n \mathcal{T}_{\text{R}_2}, n
   \mathcal{T}_{\text{R}_1}+n \mathcal{T}_{\text{R}_2},n,\gamma \big),
\end{equation}
where $\left\lfloor t/\mathcal{T}_\text{R}\right\rfloor$ is a floor function that returns an integer value that counts the 
number of full completed periods in the right bath oscillation at time $t$
with 
\begin{align}
\label{eq:f1}
\nonumber f_1\big(t_2, t_1, n,\gamma \big) &= \frac{e^{2 t_1 \gamma } (\omega_{\text{R}_1} \cos (\omega_{\text{R}_1} (t_1-n \mathcal{T}_{\text{R}_2}))-2 \gamma \sin (\omega_{\text{R}_1} (t_1-n \mathcal{T}_{\text{R}_2})))}{4 \gamma^2+\omega_{\text{R}_1}^2} \\
   &\quad + \frac{e^{2 t_2 \gamma } (2 \gamma  \sin (\omega_{\text{R}_1} ( t_2-n \mathcal{T}_{\text{R}_2}))-\omega_{\text{R}_1}
   \cos (\omega_{\text{R}_1} (t_2-n \mathcal{T}_{\text{R}_2})))}{4 \gamma^2+\omega_{\text{R}_1}^2}.
\end{align}
The second term, $I_2 (t,\gamma)$, accounts for time segments in completed right bath oscillation periods when the temperature is oscillating at frequency  $\omega_{\text{R}_2}$ and is given by 
\begin{equation}
I_2 (t,\gamma) = \sum _{n=0}^{\left\lfloor t/\mathcal{T}_\text{R}\right\rfloor -1} f_2\big((n+1) \mathcal{T}_{\text{R}_1}+ (n+1) \mathcal{T}_{\text{R}_2}, (n+1)
   \mathcal{T}_{\text{R}_1}+n \mathcal{T}_{\text{R}_2},n,\gamma\big),
\end{equation}
with 
\begin{align}
\label{eq:f2}
\nonumber f_2\big(t_2, t_1, n, \gamma \big) &= \frac{e^{2 t_1 \gamma } (\omega_{\text{R}_2} \cos (\omega_{\text{R}_2} (t_1-n \mathcal{T}_{\text{R}_1})+\phi)-2 \gamma \sin (\omega_{\text{R}_2} (t_1-n \mathcal{T}_{\text{R}_1})+\phi))}{4 \gamma^2+\omega_{\text{R}_2}^2} \\
   &\quad + \frac{e^{2 t_2 \gamma } (2 \gamma \sin (\omega_{\text{R}_2} (t_2-n \mathcal{T}_{\text{R}_1})+\phi)-\omega_{\text{R}_2}
   \cos (\omega_{\text{R}_2} (t_2-n \mathcal{T}_{\text{R}_1})+\phi))}{4 \gamma^2+\omega_{\text{R}_2}^2}.
\end{align}
The third term, $I_3 (t,\gamma)$, accounts for tail at the end of the interval $[0 , t]$ that is not included 
in the other two terms because it does not involve a full complete period of right bath oscillation.
Defining the total number of completed oscillations of the right bath at time $t$ as $n_\text{tot} = \left\lfloor t/\mathcal{T}_\text{R}\right\rfloor$, this term can be expressed as
\begin{equation}
I_3 (t,\gamma) = \begin{cases}
 f_1(t,n_\text{tot}  \mathcal{T}_{\text{R}_1}+n_\text{tot} \mathcal{T}_{\text{R}_2},n_\text{tot},\gamma), & t_\text{mod}<\mathcal{T}_{\text{R}_1}, \\
 f_1((n_\text{tot}+1) \mathcal{T}_{\text{R}_1}+n_\text{tot} \mathcal{T}_{\text{R}_2}, n_\text{tot} \mathcal{T}_{\text{R}_1}+n_\text{tot}
   \mathcal{T}_{\text{R}_2},n_\text{tot},\gamma) \\+f_2(t,(n_\text{tot}+1) \mathcal{T}_{\text{R}_1}+n_\text{tot} \mathcal{T}_{\text{R}_2},n_\text{tot},\gamma), & \mathcal{T}_{\text{R}_1} \leq t_\text{mod}<\mathcal{T}_{\text{R}}.
\end{cases}
\end{equation}
Using the expression for each part in $I(t,\gamma)$, the energy flux terms for each bath and the system can be evaluated. 
We are primarily interested in the long-time ($t \to \infty$) limit in which the system reaches a TDNES.
\EDITS{Taking the $t \to \infty$ limit after combining Eqs.~(\ref{eq:heatcurrentbathL})-(\ref{eq:heatcurrentbathS}) with Eqs.~(\ref{eq:velnoisecorrL})-(\ref{eq:velnoisecorr}) and Eq.~(\ref{eq:vsqrBrownian}) yields the  the energy fluxes in the TDNES:}
\begin{align}
\label{eq:JLlong}
J_\text{L}(t) &= 
\gamma_\text{L} k_\text{B} \bigg( T - T_\text{L} (t) 
+ \frac{2\gamma_\text{L} \Delta T_\text{L} (2 \gamma \sin(\omega_\text{L} t) - \omega_\text{L} \cos(\omega_\text{L} t))}{4 \gamma^2+\omega^2_\text{L}} 
+ 2e^{-2 \gamma t}\gamma_\text{R} \Delta T_\text{R} I(t,\gamma) 
 \bigg), \\[1ex]
\label{eq:JRlong}
J_\text{R}(t) &=  
\gamma_\text{R} k_\text{B} \bigg( T -T_\text{R} (t) + \frac{2\gamma_\text{L} \Delta T_\text{L} (2 \gamma \sin(\omega_\text{L} t) - \omega_\text{L} \cos(\omega_\text{L} t))}{4 \gamma^2+\omega^2_\text{L}} 
+ 2e^{-2 \gamma t} \gamma_\text{R} \Delta T_\text{R} I(t,\gamma) 
 \bigg),\\[1ex]
 \label{eq:JSlong}
 J_\text{sys}(t) &=  -\big(J_\text{L}(t)+ J_\text{R}(t)\big),
\end{align}
\EDITS{where the terms containing $e^{-2 \gamma t}$ do not vanish in this limit because they are multiplied by exponential terms like $e^{2 t_1 \gamma }$ and $e^{2 t_2 \gamma}$ in the function $I(t,\gamma)$ (see Eqs.~(\ref{eq:f1}) and (\ref{eq:f2})). }

\begin{figure}[]
\includegraphics[width = 9.5cm,clip]{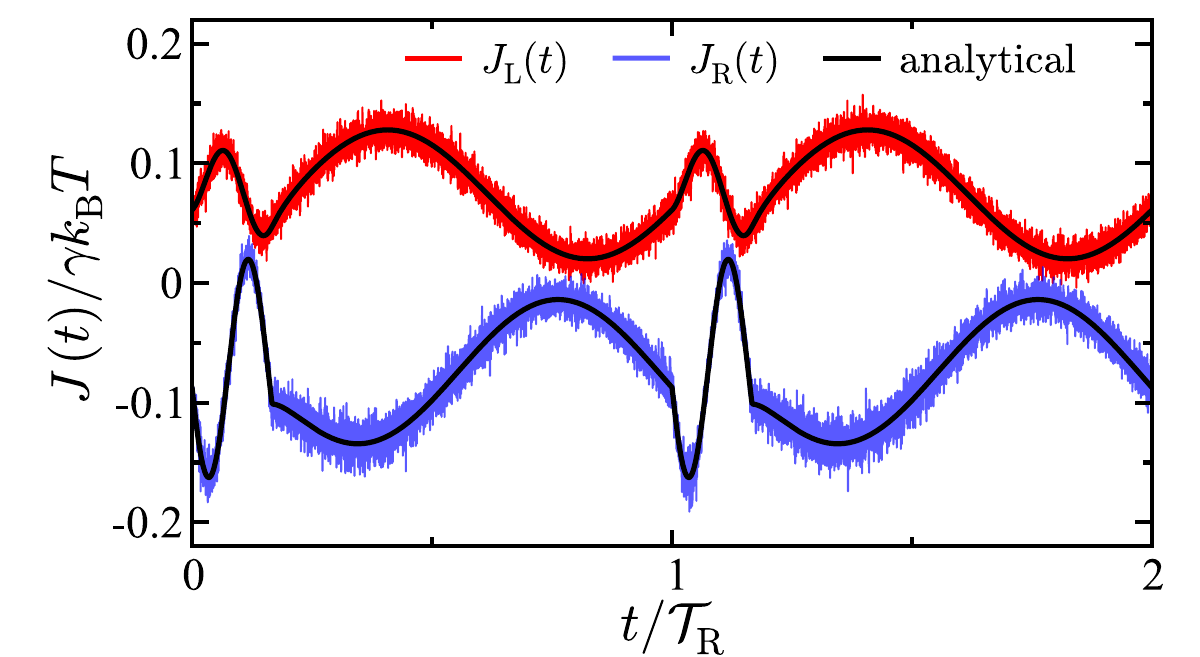}
\caption{\label{fig:flux_free}
Energy flux as a function of time for the left bath (red) and right bath (blue) for the case of a free particle.
The noisy colored curves are the results of simulations and the solid black curves 
are the analytical results. 
The parameters in the model are $\gamma = 2$ ($\gamma_\text{L} = 1$, $\gamma_\text{R} = 1$), $T^{(0)}_\text{L} = 4/5$, $T^{(0)}_\text{R} = 16/15$,
$\Delta T_\text{L} = 0$, $\Delta T_\text{R} = 0.2$, $\omega_\text{L} =0$, $\omega_{\text{R}_1} = 5$, $\omega_{\text{R}_2} = 1$, $v_0 =1$.
}
\end{figure}

Figure~\ref{fig:flux_free} shows a comparison between the theoretical results for the energy fluxes of the two baths and the corresponding results generated using molecular dynamics simulations of the system.
\EDITS{The energy flux results from the molecular dynamics simulations are generated by applying the stochastic energetics formalism  \cite{Sabhapandit2012,Sekimoto1998} to trajectories obtained by integrating Eq.~(\ref{eq:EoM1}) using the Euler-Maruyama scheme. The total number of trajectories generated was $8 \times 10^6$.}
Excellent agreement is observed between the analytical results and the simulation results.
Also observe that due to the piecewise temperature in the right bath, the energy fluxes are continuous but not smooth, having apparent derivative discontinuities at the points in time where the oscillation frequency is switched. Overall this result supports the validity of the derived theoretical expressions.

\subsection{Harmonic Oscillator}

Next, we examine the case in which the particle connecting the two heat baths moves in a harmonic potential. The general form for the derivation follows our work from Ref.~\citenum{craven2023b}, here modified to treat the frequency-shifted temperature in the right bath.  

The stochastic Langevin equation of motion for a particle in the harmonic potential $U(x) = \tfrac{1}{2} m k x^2$ is
\begin{equation}
\begin{aligned}
\label{eq:EoMHar}
\dot x &= v, \\
\dot v &= - \gamma \dot x -k x  + \xi_\text{L}(t) + \xi_\text{R}(t).
\end{aligned}
\end{equation}
The complementary equation of the stochastic ODE in Eq.~(\ref{eq:EoMHar}) is
\begin{equation}
\label{eq:eomhomohar}
		\begin{pmatrix}
			\dot{x}_\text{c}(t)\\
			\dot{v}_\text{c}(t) 
		\end{pmatrix}=
		\begin{pmatrix}
			0&1 \\
			-k &-\gamma
		\end{pmatrix}
		\begin{pmatrix}
			x(t)\\
			v(t)
		\end{pmatrix}.
\end{equation}
The fundamental matrix solution of Eq.~(\ref{eq:eomhomohar}) is
\begin{equation}
\mathbf{S}(t) = 	\left(\mathbf{v}_1 e^{\lambda_1 t} \;\; \mathbf{v}_2 e^{\lambda_2 t} \right),
\end{equation}
where $\mathbf{v}_1$ and $\mathbf{v}_2$ are eigenvectors
and 
\begin{equation}
\begin{aligned}
\lambda_1 &= \frac{1}{2}\left(-\gamma - \sqrt{\gamma^2-4 k}\right),\\
\lambda_2 &= \frac{1}{2}\left(-\gamma + \sqrt{\gamma^2-4 k}\right).
\end{aligned}
\end{equation}
Solving the initial value problem using
\begin{equation}
		\begin{pmatrix}
			x_\text{c}(t)\\
			v_\text{c}(t)
		\end{pmatrix} = \mathbf{S}(t)\mathbf{S}^{-1}(0)\begin{pmatrix}
			x_0\\
			v_0
		\end{pmatrix},
\end{equation}
allows the solution to be expressed as
\begin{equation}
\begin{pmatrix}
		x_\text{c}(t)\\
			v_\text{c}(t) 
		\end{pmatrix}	
		=  
		 \begin{pmatrix}
			\dfrac{\displaystyle e^{\lambda_1 t} \left(\lambda_2 x_0 -v_0 \right)-  e^{\lambda_2 t}\left( \lambda_1 x_0 -v_0 \right)}{\displaystyle \Delta \lambda}\\[1.5ex]
		\dfrac{\displaystyle \lambda_1 e^{\lambda_1 t}\left(\lambda_2 x_0 -v_0 \right)-\lambda_2 e^{\lambda_2 t}\left(\lambda_1 x_0 -v_0 \right) }{\displaystyle \Delta \lambda}
		\end{pmatrix},
\end{equation}
with 
\begin{equation}
\Delta \lambda = \lambda_2 - \lambda_1.
\end{equation}

With the solution to the complementary part now in hand, we can express the formal solution of the equation of motion (\ref{eq:EoMHar}) as
\begin{equation}
\begin{aligned}
\label{eq:formal}
\begin{pmatrix}
			x(t)\\
			v(t) 
		\end{pmatrix} 
		&= \mathbf{S}(t)\mathbf{S}^{-1}(0)
		\begin{pmatrix}
			x_0\\
			v_0
		\end{pmatrix} +\int_0^t \mathbf{S}(t)\mathbf{S}^{-1}(s) \begin{pmatrix}
			0\\
			\xi_\text{L}(s) +\xi_\text{R}(s)
		\end{pmatrix}\,ds,
\end{aligned}		
\end{equation}
where the first term on the RHS is the complementary  solution.
The formal solutions are
\begin{align}
\label{eq:xsolhar}
 \nonumber x(t) &=x_\text{c}(t) - \frac{1}{\Delta \lambda}\left(\int_0^t  e^{\lambda_1(t-s)}\xi_\text{L}(s)\,ds + \int_0^t  e^{\lambda_1(t-s)}\xi_\text{R}(s)\,ds\right) \\
 & \quad + \frac{1}{\Delta \lambda}\left(\int_0^t  e^{\lambda_2(t-s)}\xi_\text{L}(s)\,ds + \int_0^t  e^{\lambda_2(t-s)}\xi_\text{R}(s)\,ds\right) ,
\\[1ex]
\label{eq:vsolhar}
 \nonumber v(t) &= v_\text{c}(t)  - \frac{\lambda_1}{\Delta \lambda}\left(\int_0^t  e^{\lambda_1(t-s)}\xi_\text{L}(s)\,ds  + \int_0^t  e^{\lambda_1(t-s)}\xi_\text{R} (s)\,ds\right) \\
 &\quad + \frac{\lambda_2}{\Delta \lambda}\left(\int_0^t  e^{\lambda_2(t-s)}\xi_\text{L}(s)\,ds + \int_0^t  e^{\lambda_2(t-s)}\xi_\text{R}(s)\,ds\right).
\end{align}
Using these formal solutions, the correlation functions needed to generate expressions for the energy fluxes can be evaluated. 

The noise-velocity correlation functions are
\begin{align}
\label{eq:velnoisecorrLhar}
\nonumber\big\langle \xi_\text{L}(t) v(t)\big\rangle &= 
- \frac{\lambda_1}{\Delta \lambda}\left(\int_0^t  e^{\lambda_1(t-s)}\big\langle\xi_\text{L}(t)\xi_\text{L}(s)\big\rangle\,ds 
\right) + \frac{\lambda_2}{\Delta \lambda}\left(\int_0^t  e^{\lambda_2(t-s)}\big\langle\xi_\text{L}(t)\xi_\text{L}(s)\big\rangle\,ds\right) 
\\[1ex]
& = \frac{\gamma_\text{L} k_\text{B} T_\text{L}(t)}{m},\\
\label{eq:velnoisecorrhar}
\nonumber\big\langle \xi_\text{R}(t) v(t)\big\rangle &= 
- \frac{\lambda_1}{\Delta \lambda}\left(
\int_0^t  e^{\lambda_1(t-s)}\big\langle\xi_\text{R}(t)\xi_\text{R}(s)\big\rangle\,ds\right) + \frac{\lambda_2}{\Delta \lambda}\left(\int_0^t  e^{\lambda_2(t-s)}\big\langle\xi_\text{R}(t)\xi_\text{R}(s)\big\rangle\,ds\right)\\[1ex]
& = \frac{\gamma_\text{L} k_\text{B} T_\text{R}(t)}{m},
\end{align} 
where, for brevity, we have removed all the terms on the RHS of the correlation function that evaluate to zero.
These are the same functions derived for the free particle. Also note that they are valid for any general time-dependent form for the bath temperature.

The velocity correlation function $\big\langle v^2(t)\big\rangle$ can be evaluated by squaring the formal solution in Eq.~(\ref{eq:vsolhar}) and applying the noise correlations from Eq.~(\ref{eq:noise}) leading to
\begin{equation}
\begin{aligned}
\label{eq:vsqrHar}
\big\langle v^2(t)\big\rangle  &=  
 v^2_\text{c}(t) 
 + \left(\frac{1}{\Delta \lambda}\right)^2\!\!\Bigg(\lambda^2_1\!\!\int_0^t\!\!\int_0^t  e^{\lambda_1(2t-s_1-s_2)}\big\langle\xi_\text{L}(s_1)\xi_\text{L}(s_2)\big\rangle\,ds_1\,ds_2 \\[1ex]
&\qquad \qquad \qquad \qquad + \lambda^2_1\!\!\int_0^t\!\!\int_0^t  e^{\lambda_1(2t-s_1-s_2)}\big\langle\xi_\text{R}(s_1)\xi_\text{R}(s_2)\big\rangle\,ds_1\,ds_2 \\[1ex]
&  \qquad \qquad \qquad \qquad + \lambda^2_2\int_0^t\!\!\int_0^t  e^{\lambda_2(2t-s_1-s_2)}\big\langle\xi_\text{L}(s_1)\xi_\text{L}(s_2)\big\rangle\,ds_1\,ds_2 \\[1ex]
&\qquad  \qquad \qquad \qquad + \lambda^2_2\int_0^t\!\!\int_0^t  e^{\lambda_2(2t-s_1-s_2)}\big\langle\xi_\text{R}(s_1)\xi_\text{R}(s_2)\big\rangle\,ds_1\,ds_2 \\[1ex] 
& \qquad \qquad \qquad \qquad - 2 k\int_0^t\!\!\int_0^t   e^{\lambda_1(t-s_1)+ \lambda_2(t-s_2)}\big\langle\xi_\text{L}(s_1)\xi_\text{L}(s_2)\big\rangle\,ds_1\,ds_2 \\[1ex]
& \qquad \qquad \qquad \qquad - 2 k\int_0^t\!\!\int_0^t   e^{\lambda_1(t-s_1)+ \lambda_2(t-s_2)}\big\langle\xi_\text{R}(s_1)\xi_\text{R}(s_2)\big\rangle\,ds_1\,ds_2 \Bigg)
\\[1ex]
&=  v^2_\text{c}(t)
- \left(\frac{1}{\Delta \lambda}\right)^2\!\!\Bigg(\frac{ \gamma k_\text{B} T}{m}\bigg(\lambda_1 \left(1-e^{2 \lambda_1 t}\right) +  \lambda_2\left(1-e^{2 \lambda_2 t}\right)
+ \frac{4 k(1- e^{- \gamma t})}{\gamma}\bigg)
\\[1ex]
& \quad  
+ \frac{2 k_\text{B} \lambda^2_1}{m}\bigg(\frac{ \gamma_\text{L} \Delta T_\text{L} ( 2 \lambda_1 \sin(\omega_\text{L} t) + \omega_\text{L} \cos(\omega_\text{L} t)- \omega_\text{L} e^{2 \lambda_1 t})}{4 \lambda_1^2+\omega^2_\text{L}} 
- e^{2 \lambda_1 t} \gamma_\text{R} \Delta T_\text{R} I(t,-\lambda_1)\bigg)\\[1ex]
& \quad 
+ \frac{2 k_\text{B} \lambda^2_2}{m}\bigg(\frac{ \gamma_\text{L} \Delta T_\text{L} (2 \lambda_2 \sin(\omega_\text{L} t) + \omega_\text{L} \cos(\omega_\text{L} t) - \omega_\text{L} e^{2 \lambda_2 t})}{4 \lambda_2^2+\omega^2_\text{L}} 
- e^{2 \lambda_2 t} \gamma_\text{R} \Delta T_\text{R} I(t,-\lambda_2)\bigg) \\[1ex]
& \quad
- k\frac{4 k_\text{B}}{m}\bigg(\frac{ \gamma_\text{L} \Delta T_\text{L} (\omega_\text{L} \cos(\omega_\text{L} t)-\gamma \sin(\omega_\text{L} t) - \omega_\text{L} e^{- \gamma t})}{\gamma^2+\omega^2_\text{L}}
-e^{ -\gamma t} \gamma_\text{R} \Delta T_\text{R} P(t,\gamma)
\bigg)\Bigg),
\end{aligned}
\end{equation}
where
\begin{equation}
P(t,\gamma) = P_1 (t,\gamma) + P_2 (t,\gamma) + P_3 (t,\gamma),
\end{equation}
is another piecewise integral (representing the last integral in Eq.(\ref{eq:vsqrHar})) that is broken into three parts, each representing a different regime in the time interval $[0,t]$. This is similar to the $I(t,\gamma)$ integral calculated before. 
The first term, $P_1 (t,\gamma)$, accounts for time segments in completed right bath oscillation periods $\mathcal{T}_\text{R}$  when the temperature is oscillating at frequency $\omega_{\text{R}_1}$:
\begin{equation}
P_1 (t,\gamma) = \sum _{n=0}^{\left\lfloor t/\mathcal{T}_\text{R}\right\rfloor -1 } F_1\big((n+1) \mathcal{T}_{\text{R}_1}+ n \mathcal{T}_{\text{R}_2}, n
   \mathcal{T}_{\text{R}_1}+n \mathcal{T}_{\text{R}_2},n,\gamma \big),
\end{equation}
with 
\begin{align}
\nonumber F_1\big(t_2, t_1, n,\gamma \big) &= \frac{e^{t_1 \gamma } (\omega_{\text{R}_1} \cos (\omega_{\text{R}_1} (t_1-n \mathcal{T}_{\text{R}_2}))- \gamma \sin (\omega_{\text{R}_1} (t_1-n \mathcal{T}_{\text{R}_2})))}{ \gamma^2+\omega_{\text{R}_1}^2} \\
   &\quad + \frac{e^{ t_2 \gamma } ( \gamma  \sin (\omega_{\text{R}_1} ( t_2-n \mathcal{T}_{\text{R}_2}))-\omega_{\text{R}_1}
   \cos (\omega_{\text{R}_1} (t_2-n \mathcal{T}_{\text{R}_2})))}{ \gamma^2+\omega_{\text{R}_1}^2}.
\end{align}
The second term, $P_2 (t,\gamma)$, accounts for time segments in completed right bath oscillation periods when the temperature is oscillating at frequency  
$\omega_{\text{R}_2}$ and is given by 
\begin{equation}
P_2 (t,\gamma) = \sum _{n=0}^{\left\lfloor t/\mathcal{T}_\text{R}\right\rfloor -1} F_2\big((n+1) \mathcal{T}_{\text{R}_1}+ (n+1) \mathcal{T}_{\text{R}_2}, (n+1)
   \mathcal{T}_{\text{R}_1}+n \mathcal{T}_{\text{R}_2},n,\gamma\big),
\end{equation}
with 
\begin{align}
\nonumber F_2\big(t_2, t_1, n, \gamma \big) &= \frac{e^{t_1 \gamma } (\omega_{\text{R}_2} \cos (\omega_{\text{R}_2} (t_1-n \mathcal{T}_{\text{R}_1})+\phi)- \gamma \sin (\omega_{\text{R}_2} (t_1-n \mathcal{T}_{\text{R}_1})+\phi))}{\gamma^2+\omega_{\text{R}_2}^2} \\
   &\quad + \frac{e^{ t_2 \gamma } ( \gamma \sin (\omega_{\text{R}_2} (t_2-n \mathcal{T}_{\text{R}_1})+\phi)-\omega_{\text{R}_2}
   \cos (\omega_{\text{R}_2} (t_2-n \mathcal{T}_{\text{R}_1})+\phi))}{\gamma^2+\omega_{\text{R}_2}^2}.
\end{align}
The third term, $P_3 (t,\gamma)$, accounts for tail at the end of the interval $[0 , t]$ that is not included 
in the other two terms because it does not involve a full complete period of right bath oscillation.
This term can be expressed as
\begin{equation}
P_3 (t,\gamma) = \begin{cases}
 F_1(t,n_\text{tot}  \mathcal{T}_{\text{R}_1}+n_\text{tot} \mathcal{T}_{\text{R}_2},n_\text{tot},\gamma), & t_\text{mod}<\mathcal{T}_{\text{R}_1}, \\
 F_1((n_\text{tot}+1) \mathcal{T}_{\text{R}_1}+n_\text{tot} \mathcal{T}_{\text{R}_2}, n_\text{tot} \mathcal{T}_{\text{R}_1}+n_\text{tot}
   \mathcal{T}_{\text{R}_2},n_\text{tot},\gamma) \\
   +F_2(t,(n_\text{tot}+1) \mathcal{T}_{\text{R}_1}+n_\text{tot} \mathcal{T}_{\text{R}_2},n_\text{tot},\gamma), & \mathcal{T}_{\text{R}_1} \leq t_\text{mod}<\mathcal{T}_{\text{R}},
\end{cases}
\end{equation}
where $n_\text{tot} = \left\lfloor t/\mathcal{T}_\text{R}\right\rfloor$ as before.
The energy fluxes for the harmonic potential (the left bath, right bath, and system energy fluxes) can now be readily obtained. 
In the long-time limit, the energy flux expressions in the TDNES are:
\begin{align}
\label{eq:JLlonghar}
\nonumber J_\text{L}(t) &= - \gamma_\text{L} k_\text{B} T_\text{L}(t) - \gamma_\text{L} \left(\frac{1}{\Delta \lambda}\right)^2\!\!\Bigg(k_\text{B} T \big(
4 k-\gamma^2\big)
\\[1ex]
\nonumber & \quad  
+ 2 k_\text{B} \lambda^2_1\bigg(\frac{ \gamma_\text{L} \Delta T_\text{L} ( 2 \lambda_1 \sin(\omega_\text{L} t) + \omega_\text{L} \cos(\omega_\text{L} t))}{4 \lambda_1^2+\omega^2_\text{L}} 
- e^{2 \lambda_1 t} \gamma_\text{R} \Delta T_\text{R} I(t,-\lambda_1)\bigg)\\[1ex]
\nonumber & \quad 
+ 2 k_\text{B} \lambda^2_2\bigg(\frac{ \gamma_\text{L} \Delta T_\text{L} (2 \lambda_2 \sin(\omega_\text{L} t) + \omega_\text{L} \cos(\omega_\text{L} t))}{4 \lambda_2^2+\omega^2_\text{L}} 
- e^{2 \lambda_2 t} \gamma_\text{R} \Delta T_\text{R} I(t,-\lambda_2)\bigg) \\[1ex]
& \quad
- 4 \lambda_1 \lambda_2 k_\text{B}\bigg(\frac{ \gamma_\text{L} \Delta T_\text{L} (\omega_\text{L} \cos(\omega_\text{L} t)-\gamma \sin(\omega_\text{L} t))}{\gamma^2+\omega^2_\text{L}} 
-e^{ -\gamma t} \gamma_\text{R} \Delta T_\text{R} P(t,\gamma)\bigg)\Bigg), \\[1ex]
\label{eq:JRlonghar}
\nonumber J_\text{R}(t) &= - \gamma_\text{R} k_\text{B} T_\text{R}(t) - \gamma_\text{R} \left(\frac{1}{\Delta \lambda}\right)^2\!\!\Bigg(k_\text{B} T \big(
4 k-\gamma^2\big)
\\[1ex]
\nonumber & \quad  
+  2 k_\text{B} \lambda^2_1\bigg(\frac{ \gamma_\text{L} \Delta T_\text{L} ( 2 \lambda_1 \sin(\omega_\text{L} t) + \omega_\text{L} \cos(\omega_\text{L} t))}{4 \lambda_1^2+\omega^2_\text{L}} 
- e^{2 \lambda_1 t} \gamma_\text{R} \Delta T_\text{R} I(t,-\lambda_1)\bigg)\\[1ex]
\nonumber & \quad 
+ 2 k_\text{B} \lambda^2_2\bigg(\frac{ \gamma_\text{L} \Delta T_\text{L} (2 \lambda_2 \sin(\omega_\text{L} t) + \omega_\text{L} \cos(\omega_\text{L} t))}{4 \lambda_2^2+\omega^2_\text{L}} 
- e^{2 \lambda_2 t} \gamma_\text{R} \Delta T_\text{R} I(t,-\lambda_2) \bigg) \\[1ex]
& \quad
- 4 \lambda_1 \lambda_2 k_\text{B}\bigg(\frac{ \gamma_\text{L} \Delta T_\text{L} (\omega_\text{L} \cos(\omega_\text{L} t)-\gamma \sin(\omega_\text{L} t))}{\gamma^2+\omega^2_\text{L}} 
-e^{ -\gamma t} \gamma_\text{R} \Delta T_\text{R} P(t,\gamma)\bigg)\Bigg), \\[1ex]
\label{eq:JSlonghar}
\nonumber J_\text{sys}(t) &=  -\big(J_\text{L}(t)+ J_\text{R}(t)\big).
\end{align}
These energy flux expressions are the primary theoretical results presented in this article. They can be used to examine the hysteresis behavior in the model 
and to explore how different sets of parameters and system conditions affect the hysteresis curves.

\section{Heat Transport Hysteresis \label{sec:hys}}

\begin{figure}[t]
\includegraphics[width = 9.5cm,clip]{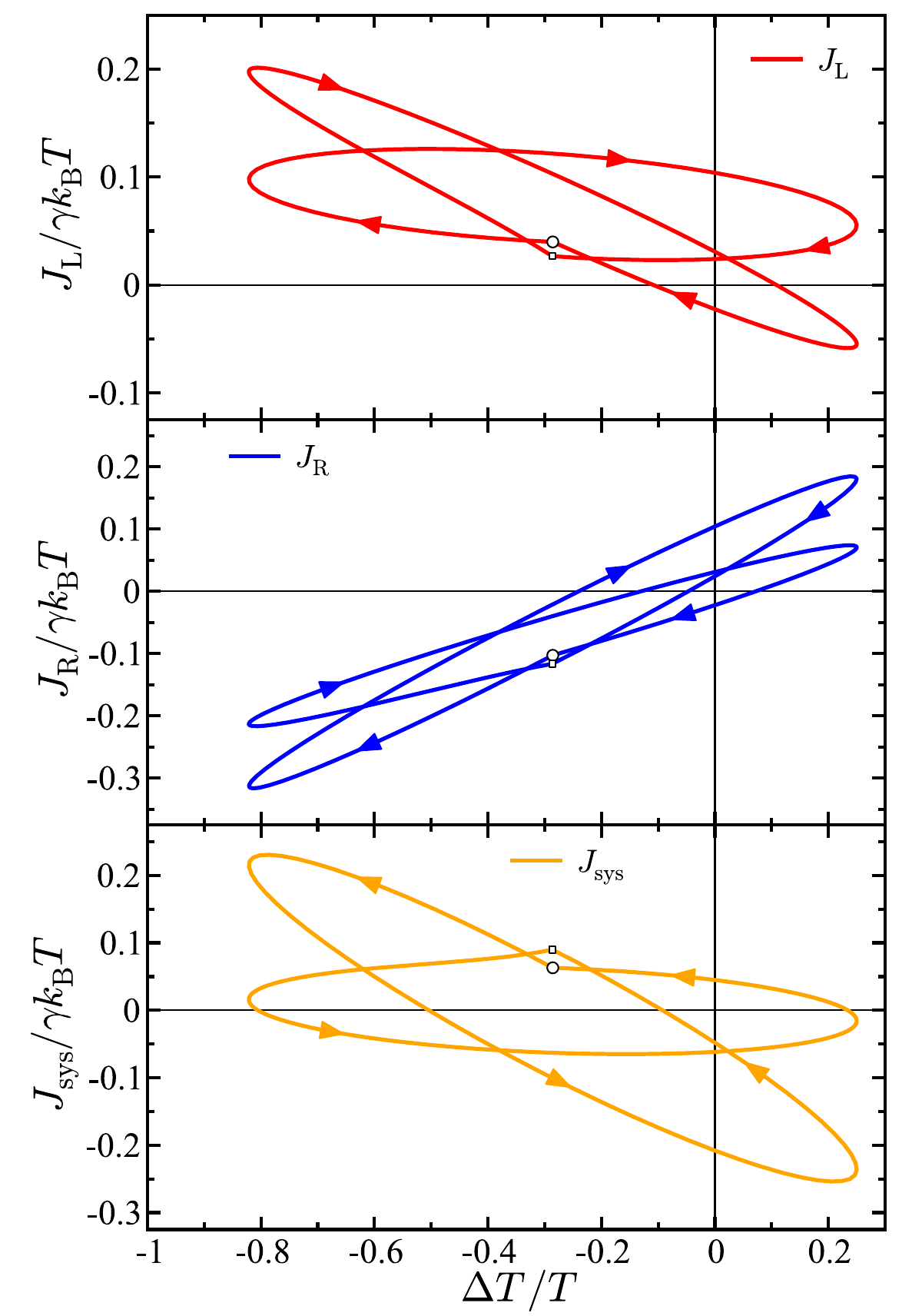}
\caption{\label{fig:Hys_1}
Energy flux as a function of temperature difference for the left bath (top), right bath (middle), and the system (bottom) which is in this case a free particle.
\EDITS{The larger circular marker on each curve marks the value of the respective energy flux at beginning of the overall period.
The smaller square marker on each curve marks the point where the frequency switches.}
The parameters in the model are $\gamma = 2$ ($\gamma_\text{L} = 1$, $\gamma_\text{R} = 1$), $T^{(0)}_\text{L} = 4/5$, $T^{(0)}_\text{R} = 16/15$,
$\Delta T_\text{L} = 0$, $\Delta T_\text{R} = 0.5$, $\omega_\text{L} =0$, $\omega_{\text{R}_1} = 10$, $\omega_{\text{R}_2} = 1$.
}
\end{figure}

The time-dependence of the temperature gradient gives rise to heat transport hysteresis effects.
This means that depending on the time in the oscillation period, the same temperature difference can generate different energy flux values.
Heat transport hysteresis is observed when the energy flux generated by a temperature gradient that is oscillating in time is out of phase with the  oscillation pattern of the gradient itself.
Figure~\ref{fig:Hys_1} shows the energy fluxes in the model in the $J \times \Delta T$ plane where both the energy fluxes $J$ and the temperature difference $\Delta T$ depend on time. Specifically, each panel in the figure shows
the energy flux in the left bath (top), the right bath (middle), and the system (bottom) as a function of the time-dependent temperature difference between baths $\Delta T(t)$.
The starting point of each hysteresis loop is denoted by a circular marker---this is the start of the oscillation period $\mathcal{T}_\text{R}$.
The presented results in Figure~\ref{fig:Hys_1} are for the case when only the right bath is oscillating and the left bath temperature is static,
so there are two primary frequencies in the model.
Correspondingly, two prominent curves can be observed in the hysteresis curves, one for each of the two frequencies.
\EDITS{All the energy fluxes start on one of the two prominent curve shapes in the figure, then,
when the frequency is switched (shown by the knee in the curve and denoted by a small square marker), the energy flux switches to the other curve.
However, the motion is continuous and periodic overall and at the end of the total period the energy flux returns to the starting point denoted by the larger circular marker.}
This is observed in the left and right bath hysteresis curves as well as for the system.

\begin{figure}[t]
\includegraphics[width = 9.5cm,clip]{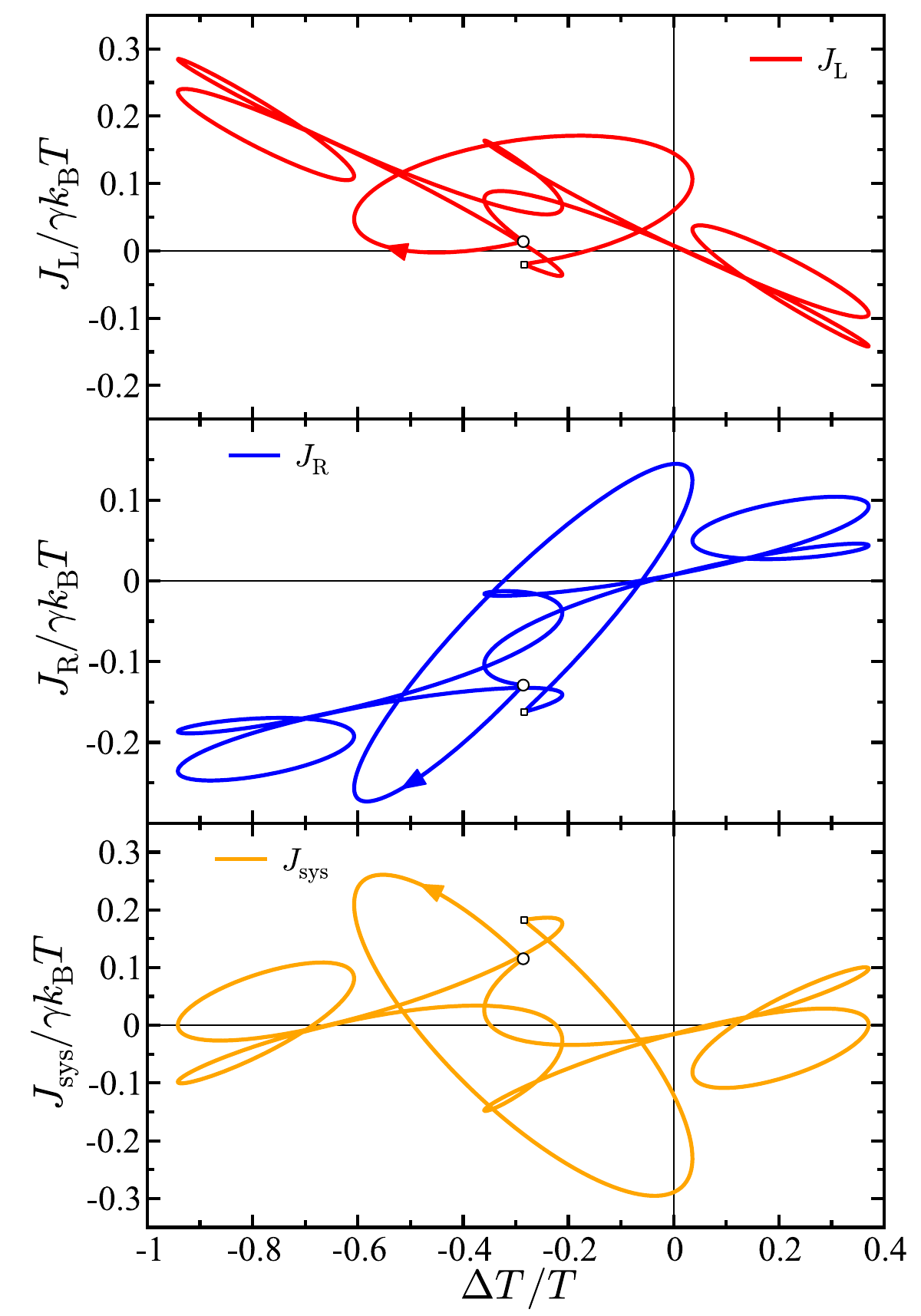}
\caption{\label{fig:Hys_2}
Energy flux as a function of temperature difference for the left bath (top), right bath (middle), and the system (bottom) which in this case is a free particle.
\EDITS{The larger circular marker on each curve marks the value of the respective energy flux at beginning of the overall period.
The smaller square marker on each curve marks the point where the frequency switches.}
The parameters in the model are $\gamma = 2$ ($\gamma_\text{L} = 1$, $\gamma_\text{R} = 1$), $T^{(0)}_\text{L} = 4/5$, $T^{(0)}_\text{R} = 16/15$,
$\Delta T_\text{L} = 0.2$, $\Delta T_\text{R} = 0.5$, $\omega_\text{L} =5$, $\omega_{\text{R}_1} = 5$, $\omega_{\text{R}_2} = 1$.
}
\end{figure}

Figure~\ref{fig:Hys_2} shows the results when the right bath is undergoing frequency switching between fast and slow regimes and 
the left bath is also oscillating. This creates a complex interplay between the different driving frequencies in the system. As a result, a complex multi-loop heat transport hysteresis pattern emerges. Again, the starting point in each hysteresis loop is denoted by a circular marker.
The hysteresis pattern begins on a simple curve, but when the frequency is shifted (at the prominent knee in the curve denoted by a small square marker), the hysteresis curve shape  changes abruptly with the energy flux changing direction in the $J \times \Delta T$ plane. 
After the frequency is shifted, the hysteresis curve has multiple loops, pinched loops, and interesting geometrical features. 
The results in Figure~\ref{fig:Hys_2} illustrate how frequency switching can be used to generate highly complex heat transport hysteresis patterns.
\EDITS{The complex shape of the curve is caused by the multiple frequencies in the system and because, in this case, the left bath temperature is also oscillating but with a different amplitude than the right bath. This causes a complex form for $\Delta T (t)$ which in turn causes the multi-loop hysteresis pattern with multiple pinched loops to appear in the $J \times \Delta T$ plane. }

\begin{figure}[t]
\includegraphics[width = 9.5cm,clip]{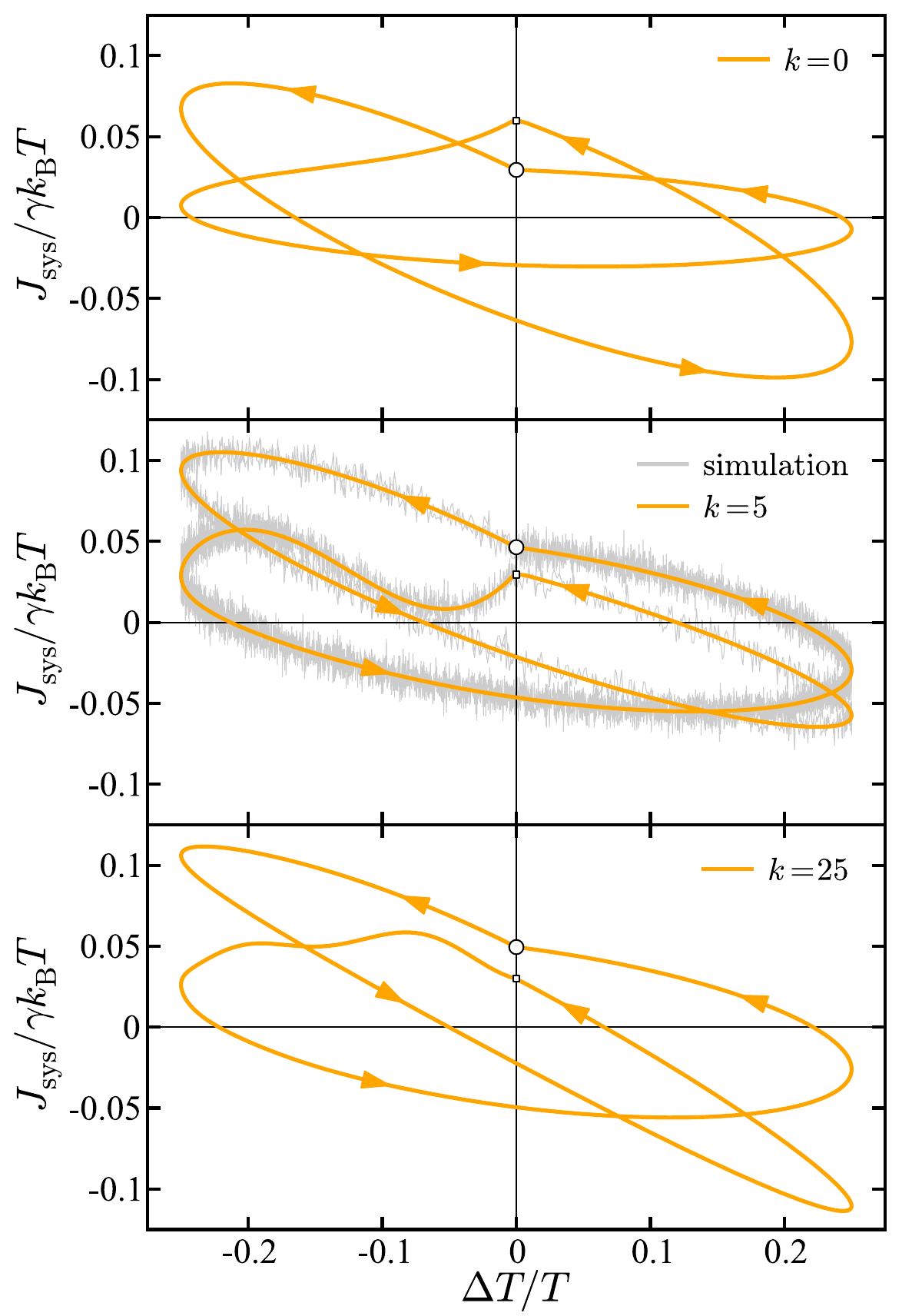}
\caption{\label{fig:Hys_1_HO}
Energy flux as a function of temperature difference for a harmonic oscillator with force constants $k=0$ (top),  $k=5$ (middle), and  $k=25$ (bottom).
\EDITS{The larger circular marker on each curve marks the value of the respective energy flux at beginning of the overall period.
The smaller square marker on each curve marks the point where the frequency switches.
The solid curves are the exact analytical results and the noisy gray curve in the $k = 5$ panel is the result from simulation.}
The parameters in the model are $\gamma = 2$ ($\gamma_\text{L} = 1$, $\gamma_\text{R} = 1$), $T^{(0)}_\text{L} = 4/5$, $T^{(0)}_\text{R} = 4/5$,
$\Delta T_\text{L} = 0.0$, $\Delta T_\text{R} = 0.2$, $\omega_\text{L} =0$, $\omega_{\text{R}_1} = 5$, $\omega_{\text{R}_2} = 1$.
}
\end{figure}

Figure~\ref{fig:Hys_1_HO} illustrates how changing the force constant in the harmonic potential can affect the hystersis curve shape.
The panels from top to bottom show the energy flux hysteresis curve for $k=0$ (top),  $k=5$ (middle), and  $k=25$ (bottom).
The primary observation is that changing the force constant gives rise to different hysteresis shapes and geometries.
As the force constant is increased, the knee in the curve (the point where the frequency shifts in the middle of the overall period, becomes
less pronounced. In fact, for $k=25$, the knee is difficult to identify signifying that the the curve is approaching smoothness.
\EDITS{The results of molecular dynamics are shown in the $k=5$ panel (the middle panel). Two observations are of note: (a) The simulation results are in excellent agreement with the analytical results suggesting the validity of the derived energy flux expressions and (b) It can be observed that the density of sampled points in the noisy simulation results is larger on some parts of the curves but smaller on others. This is because the simulation results are sampled at a fixed time interval ($\Delta t = 0.001)$ and so for the lower frequency part of the frequency-switched hysteresis curves the sampling density is higher. Then, when the frequency switches to a higher frequency, the sampling density becomes lower.}
It is also interesting to note by comparing each panel in the figure that varying the force constant does not just change the width and length of the curve, but changes it's overall shape,
for example, it can be observed by comparing the shape of the curves across force constants in the upper left quadrant, that the concavity of the curve 
changes as the force constant is altered. 
\EDITS{The cause of this effect is that when $k\neq 0$ another frequency is effectively introduced into the system. This frequency is the frequency of the harmonic potential. Therefore, a complex interplay of phases, frequencies, and oscillation patterns arises between the bath oscillation frequencies, the frequency of the harmonic potential, and the frequencies characterized by the system-bath coupling terms for each bath.}

\section{Conclusions}
In summary, the presented work provides a theoretical description of the effects of frequency switching on heat transport properties and the resulting hysteresis behavior in nanoscale systems driven by an oscillating temperature gradient. By employing stochastic thermodynamics to derive expressions for the energy fluxes and corresponding hysteresis curves, we have demonstrated that frequency shifting can be applied to control energy flux and reshape hysteresis curves. The findings in this article illustrate that transitions between fast and slow oscillation regimes influence the emergence of pinched hysteresis loops and intricate multi-loop hysteresis structures. This work offers new insights into the interplay between time-dependent thermal gradients and heat transport. The presented results have implications in the development and optimization of nanoscale thermal devices, particularly in the context of thermal neuromorphic computers and advanced thermal management technologies. Future work will examine how frequency shifting can be used to design thermal memristors and memcapacitors for use in thermal neuromorphic computers. \EDITS{Experimental realizations of the phenomena discussed in this article could potentially be generated using  molecular junction set-ups \cite{Reddy2007,Tan2011,Lee2013,Kim2014,Venkataraman2015,Reddy2019nature,Mosso2019, Zimbovskaya2020} in which a molecule bridge (the system) connects two electrodes (the heat baths). Generating and then controlling a temperature bias across the molecular junction will induce an energy flux through the molecular bridge, and oscillating that temperature bias over time could result in the heat transport hysteresis effects predicted here being observed.}

\acknowledgments{This research was performed in part at the Center for Nonlinear Studies (CNLS) at Los Alamos National Laboratory (LANL). The computing resources used to perform this research were provided by the LANL Institutional Computing Program. We acknowledge support from the Los Alamos National Laboratory (LANL) 
Directed Research and Development funds (LDRD).
}

\bibliography{main.bbl}

\end{document}